\documentclass[10pt]{article}
\usepackage{amsmath,amssymb,fullpage}

\usepackage{enumerate}

\usepackage{graphicx}

\usepackage{makeidx}
\usepackage[utf8]{inputenc}
\usepackage{wrapfig}
\usepackage{amsmath}
\usepackage{amsfonts}
\usepackage{mathrsfs}
\usepackage{amssymb}
\usepackage{latexsym}
\usepackage{amsbsy}
\usepackage{color}
\usepackage{bbm}

\usepackage{mathrsfs}

\usepackage{wasysym}

\providecommand{\otherindexspace}[1]{}

\usepackage{amsthm}

\newtheorem{theorem}{Theorem}[section]

\newtheorem{proposition}[theorem]{Proposition}
\newtheorem{remark}[theorem]{Remark}

\newtheorem{definition}[theorem]{Definition}

\DeclareMathOperator{\essinf}{essinf}
\DeclareMathOperator{\esssup}{esssup}

\numberwithin{equation}{section}

\def \R{\mathbb {R}}

\def \N{\mathbb{N}}

\def \esssup{ess \sup}

\usepackage{fancyhdr}

\usepackage{graphics}
\usepackage{graphicx}

\DeclareGraphicsExtensions{.eps,.bmp,.jpg,.pdf,.mps,.png,.gif}

\makeatletter
\def\titre{\@title}
\makeatother

\title{An optional decomposition of $\mathscr{Y}^{g,\xi}-submartingales$ and applications to the hedging of American options in incomplete markets}

\author{Roxana Dumitrescu}

\begin{document}



\date{}

\maketitle

\begin{abstract}
In the recent paper \cite{DESZ}, the notion of $\mathscr{Y}^{g,\xi}$-submartingale processes has been introduced.
Within a jump-diffusion model, we prove here that a process $X$ which satisfies the simultaneous $\mathscr{Y}^{\mathbb{Q},g,\xi}$ -submartingale property under a suitable family of equivalent probability measures $\mathbb{Q}$, admits a \textit{nonlinear optional decomposition}. This is an analogous result to the well known optional decomposition of simultaneous (classical and $\mathscr{E}^g$-)supermartingales. We then apply this decomposition to the super-hedging problem of an American option in a jump-diffusion model, from the buyer's point of view. We obtain an \textit{infinitesimal characterization} of the buyer's superhedging price, this result being completely new in the literature. Indeed, it is well known that the seller's superheding price of an American option admits an infinitesimal representation in terms of the \textit{minimal supersolution of a constrained reflected BSDE}. To the best of our knowledge, no analogous result has been established for the buyer of the American option in an incomplete market. Our results fill this gap, and show that the buyer's super-hedging price admits an infinitesimal characterization in terms of  the \textit{maximal subsolution of a constrained reflected BSDE}.
\end{abstract}

\bigskip

\textbf{Keywords:} American  options, buyer's price, incomplete markets, nonlinear pricing, reflected BSDEs with constraints, nonlinear optional decomposition\\

\textbf{AMS MSC 2010:} Primary 60G40; 93E20; 60H30, Secondary  60G07; 47N10.


\section{Introduction}
 Consider a probability space $(\Omega, \mathscr{F}, \mathbb{P})$ which supports a Brownian motion $W$ and an independent Poisson Process $N$ of intensity $\lambda$. Let $(\mathscr{F}_t)_{t \in [0,T]}$ be the $\mathbb{P}$-augmented filtration generated by $W$ and $N$.

In the recent paper \cite{DESZ}, the notion of a $\mathscr{Y}^{g,\xi}$-submartingale process has been introduced, where given a right-continuous process $\xi$, a nonlinear map $g(t,\omega,y,z,k)$, a terminal time $T$ and a random variable $\zeta$, the operator $\mathscr{Y}_{\cdot, T}^{g,\xi}(\zeta)$ is defined as follows:
$$\mathscr{Y}_{t, T}^{g,\xi}(\zeta):= \esssup_{\nu \geq t} \mathscr{E}^g_{t,\nu}\left(\xi_\nu \textbf{1}_{\nu <T}+\zeta \textbf{1}_{\nu=T}\right),$$
where $\mathscr{E}^g_{\cdot, \theta}(\zeta)$ corresponds to the solution of a backward stochastic differential equation associated with nonlinear driver $g$, terminal time $\theta$ and terminal condition $\zeta$. To illustrate the main results of our paper, we consider here $g \equiv 0$ and the associated nonlinear operator $\mathscr{Y}_{t, T}^{\xi}(\zeta)$, which takes then the form:
$$\mathscr{Y}_{t, T}^{\xi}(\zeta):= \esssup_{\nu \geq t} \mathbb{E}\left[\xi_\nu \textbf{1}_{\nu <T}+\zeta \textbf{1}_{\nu=T}| \mathscr{F}_t \right].$$

 A process $X$ is called a $\mathscr{Y}^{\xi}$-submartingale if $X_t \geq \xi_t$ for all $t$ and $X_S \leq \mathscr{Y}^{\xi}(X_{\theta})$. In \cite{DESZ}, it has been shown that, if $X$ is a right-continuous left-limited $\mathscr{Y}^{\xi}$-submartingale, then it admits a \textit{Doob-Meyer-Mertens} decomposition which can be written as follows:
\begin{align}
    -dX_t=-Z_tdW_t - U_t d \Tilde{N}_t+dA_t-dA'_t;\\
    X_t \geq \xi_t;\\
    \int_0^T (X_{s^-}-\xi_{s^-})dA_s=0,
\end{align}
with $A$ and $A'$ increasing right-continuous left-limited predictable processes such that $A_0=A'_0=0$.\\

We place ourselves in a jump-diffusion model and consider the process $S$ which follows the dynamics:
\begin{align}
    dS_t=S_{t^-}dM_t^S,
\end{align}
where $M_\cdot^S:=\int_0^\cdot \sigma_t^1 dW_t+\int_0^\cdot \sigma_t^2 d\Tilde{N}_t$, for some bounded coefficients $\sigma^1, \sigma^2$ and let $\mathcal{M}_0$ the family of equivalent martingale measures $\mathbb{Q}$ on $(\Omega, \mathscr{F}, \mathbb{P})$ for the process $S$.
Our first main contribution is to show that a process $X$, which is a $\mathscr{Y}^{\mathbb{Q},\xi}$-submartingale under each probability measure $\mathbb{Q} \in \mathcal{M}_0$, admits an \textit{optional decomposition} of the following form
\begin{align*}
    -dX_t=-Z_t\sigma_t^{-1}dM_t^S-dk'_t+dk_t; \nonumber\\
    X_t \geq \xi_t; \\
    dk'_t \perp dk_t; \nonumber\\
    \int_0^T (X_{s^-}-\xi_{s^-})dk_s=0 \text{\,\, a.s.\,\,}
\end{align*}
where the processes $k$ and $k'$ are increasing right-continuous left-limited processes such that $k_0=k'_0=0$. 

Our results are given in the general setting of an operator $\mathscr{Y}^{g,\xi}$, which leads to a \textit{nonlinear optional decomposition} for those processes which satisfy a simultaneous $\mathscr{Y}^{\mathbb{Q},g,\xi}$-submartingale property under a suitable family of equivalent probability measures. This is an analogous result to the well known \textit{optional decomposition of supermartingales} (see e.g. \cite{CviK}, \cite{KaKou}, \cite{Kramkov} in the linear case, and extended to the nonlinear setting in \cite{GQS1}).

Our second contribution is to apply this result to the problem of hedging of American options in an incomplete jump-diffusion market model, where typically the process $(S_t)$ represents the risky asset's price. This type of decomposition allows to obtain an \textit{infinitesimal characterization} of the buyer's super-hedging price  of an American option, which represents a new result in the literature. Indeed, it is well known that the seller's superheding price of an American option can be characterized as \textit{the minimal supersolution of a constrained reflected BSDE} (see e.g. Proposition 6.13 in \cite{CviK} and Theorem 4.6. in \cite{GQS2}). To the best of our knowledge, no analogous result has been established for the buyer of the American option, the roles of the buyer and of the seller being asymmetric in the context of American options in incomplete markets. Our results fill this gap, and show that the buyer's super-hedging price can be characterized as the \textit{maximal subsolution of a constrained reflected BSDE}. Furthermore, our Theorem \ref{buyamericano} provides a \textit{dynamic} version
of Theorem 5.13 in \cite{KaKou} in the case of a linear market, extended using a similar approach to the nonlinear case in \cite{GQS2}, Theorem 7.12. These results from the previous literature give a pricing-hedging duality result for the lower bound
of the arbitrage-free option prices only at time $t = 0$.

The paper is organized as follows: in Section 2, we introduce the financial market model. In Section 3, we present the main results, in particular the \textit{nonlinear optional decomposition} of processes which are $\mathscr{Y}^{\mathbb{Q},g,\xi}$-submartingales under a suitable family of equivalent probability measures $\mathbb{Q} \in \mathcal{M}_0$ and the application to the pricing of American options in incomplete markets (i.e. the infinitesimal characterization of of the buyer's price process in terms of the \textit{maximal subsolution of a reflected BSDE}). In Section 4, we collect the proofs.\\

\noindent \textbf{Notations and definitions.} Let $(\Omega, \mathscr{F}, \mathbb{P})$ be a complete probability space, which supports a one-dimensional standard Brownian motion $W$ and an independent Poisson process with intensity $\lambda>0$. We denote by $\mathbb{F}:=\{\mathscr{F}_t, \, t \geq 0\}$ the $\mathbb{P}$-\textit{augmentation} of the filtration generated by $W$ and $N$.  In the paper, $\Tilde{N}$ represents the compensated Poisson process and it is given by $\Tilde{N}_t:=N_t-\lambda t$.\\
\noindent Let $T>0$. All processes encountered throughout the paper will be defined on the fixed, finite horizon $[0,T]$. We introduce the following sets:
\begin{itemize}
\item ${\mathbf S}^{2}$ 
is the set of ${\mathbb F}$-optional processes $\varphi$ such that $\mathbb{E}[ess \sup_{ \tau \in \mathcal{T}} |\varphi_\tau | ^2] < +\infty$.
\item ${\mathbf A}^2$  is the set of real-valued non decreasing RCLL predictable
 processes $A$ with $A_0 = 0$ and $\mathbb{E}(A^2_T) < \infty$.

\item ${\mathbf A}_{o}^{2}$  is the set of real-valued non decreasing RCLL optional
 processes $A$ with $A_0 = 0$ and $\mathbb{E}(A^2_T) < \infty$.
 
 \item  ${\mathbf C}^2$  is the set of real-valued purely discontinuous non decreasing RCLL optional
 processes $C$ with $C_{0} = 0$ and $\mathbb{E}(C^2_T) < \infty$.

\item ${\mathbf H}^2$  is the set of ${\mathbb  F}$-predictable processes $Z$ such that
 $
 \| Z\|^2_{\mathbf{H}^2}:= \mathbb{E}\Big[\int_0^T|Z_t|^2dt\Big]<\infty \,.
 $
\end{itemize}
Moreover,  $\mathcal{T}_0$ denotes the set of stopping times $\tau$ such that $\tau \in [0,T]$ a.s.\, and for each $S$ in $\mathcal{T}_0$, $\mathcal{T}_{S}$ is  the set of stopping times $\tau$ such that $S \leq \tau \leq T$ a.s.\\

\noindent \noindent We now give the definition of an-\textit{admissible driver}.

\begin{definition}[\textbf{{\em Admissible} driver}]\label{defd}
A  function $g$
is said to be a {\em driver} if\\
$g: \Omega \times [0,T] \times \R^3  \rightarrow \R $; 
$(\omega, t,y, z, k) \mapsto  g(\omega, t,y,z,k) $
which   is $ {\mathcal P} \otimes {\mathcal B}(\R^3) 
- $ measurable, and such that
 $g(.,0,0,0) \in {\mathbf{H}}^2$.
 
A driver $g$ is called an {\em admissible driver} if moreover there exists a constant $ C \geq 0$ such that 
$dP \otimes dt$-a.s.\,,
for each  $(y, z, k)$, $(y_1, z_1, k_1)$, $(y_2, z_2, k_2)$,
\begin{equation}\label{lip}
|g(\omega, t, y, z_1, k_1) - g(\omega, t, y, z_2, k_2)| \leq
C ( |y_1-y_2|+|z_1 - z_2| + |k_1 - k_2 |).
\end{equation}
\end{definition}

Let us now recall the definition of the $\mathscr{E}^g$-conditional expectation operator and of a  $\mathscr{E}^g$-submartingale process.

\begin{definition}[{\bf Nonlinear operator ${\mathscr{E}^{g}}$}] 
The $g$-conditional expectation, denoted by $\mathscr{E}^g$, is the operator defined for each $T' \in [0,T]$ and for each $\zeta \in \textbf{L}(\mathscr{F}_{T'})$ by $\mathscr{E}^f_{t, T'}(\zeta):=X_t$, for each $t \in [0,T]$, where $(X,Z,U) \in \textbf{S}^2 \times (\textbf{H}^2)^2$ is the solution of the BSDE associated with driver $f$, terminal time $T'$ and terminal condition $\zeta$ and driven by $W$ and $\Tilde{N}$, that is
\begin{align}
    -dX_t=g(t,X_t,Z_t, U_t)dt-Z_t dW_t-U_td\Tilde{N}_t.
\end{align}
\end{definition}

\begin{definition}[{\bf ${\mathscr{E}^{g}}$-martingale process}]
An optional process $X \in \mathbf{S}^2$ is said to be a strong  $\mathscr{E}^{g}$-\emph{martingale process} if, for all $S, S^{'}$ $ \in \mathcal{T}_0$ such that $S \geq S^{'}$ a.s., 
\begin{eqnarray*} 
\mathscr{E}^{g}_{S',S} (X_S) =  X_{S^{'}} \quad
\,\mbox{a.s.}
\end{eqnarray*}
\end{definition}

Let $\xi \in \textbf{S}^2$ be a right-continuous process. We recall now the definition of the nonlinear $\mathscr{Y}^{g,\xi}$-operator.

\begin{definition}[{\bf Nonlinear operator ${\mathscr{Y}^{g,\xi}}$}]  

For each $\tau \in \mathcal{T}_0$ and each $\zeta \in L^2(\mathcal{F}_\tau)$   such that $\zeta \geq \xi_\tau$ a.s., we define $\mathscr{Y}^{g,\xi}_{\cdot,\tau}(\zeta):=Y_\cdot$, where $Y_\cdot$ corresponds to the first componant of the solution of the reflected BSDE associated with terminal time $\tau$, driver $g$ and lower obstacle $(\xi_t \textbf{1}_{t<\tau}+\zeta \textbf{1}_{t \geq \tau})$.
\end{definition}

We finally present the notion of a strong $\mathscr{Y}^{g,\xi}$-submartingale process, which has been recently introduced in \cite{DESZ}. 
\begin{definition}[{\bf ${\mathscr{Y}^{g,\xi}}$-submartingale process}]
An optional process $X $ is said to be a strong  $\mathscr{Y}^{g,\xi}$-\emph{submartingale process} (resp. a strong $\mathscr{Y}^{g,\xi}$- \emph{martingale process}) if for each $S\in \mathcal{T}_{0}$, $X_S \geq \xi_S$ and if, for all 
$S, S^{'}$ $ \in \mathcal{T}_0$ such that $S \geq S^{'}$ a.s., 
\begin{eqnarray*} 
\mathscr{Y}^{g,\xi}_{S',S} (X_S)  \geq  X_{S^{'}} \quad
\,\mbox{a.s.,} && {\rm (resp.} \quad  \mathscr{Y}^{g,\xi}_{S',S} (X_S)  =  X_{S^{'}} \quad
\,\mbox{a.s.).}
\end{eqnarray*}
\end{definition}

\

\section{The model}
We place ourselves directly in the financial market model $\mathscr{M}$ that consists of one risk-free asset whose price process $S^0=(S_t^0)_{0 \leq t \leq T}$ satisfies
\begin{align}
dS_t^0=S_t^0 r_tdt
\end{align}
 and one risky asset with price process $S=(S_t)_{0 \leq t \leq T}$ which evolves according to the equation
\begin{align}\label{one}
dS_t=S_{t^-}\bigg(\mu_t dt+\sigma^1_t dW_t+\sigma^2_t d\Tilde{N}_t\bigg).
\end{align}
\noindent The coefficients associated with the model $\mathscr{M}$, that is, the processes $r_t$, $\mu_t$, $\sigma^1_t$ and $\sigma^2_t$ are supposed to be predictable, satisfying $\sigma^1 > 0$ and  $\sigma^2> -1$, and such that $\sigma^1$, $\sigma^2$, $\sigma^{-1}$ and $\mu$ are bounded.\\


\textbf{Portfolio dynamics.} We consider an investor, whose initial wealth at time $0$ is equal to $x $ and who can invest his wealth in the two assets of the market. 
The amount invested at each time $t$ in the risky asset is denoted by $\varphi_t$.

For an initial wealth $x \in \mathbb{R}$ and a portfolio strategy $\varphi \in \mathbf{H}^2$, we denote by $V^{x, \varphi}_t$ (or, to simplify the notation, by $V_t$) the value of the associated portfolio  or {\em wealth}, which is supposed to satisfy the following dynamics:
\begin{equation}\label{weaun}
-dV_t= f(t,V_t, {\varphi_t} \sigma^1_t) dt - {\varphi_t} \sigma^1_t dW_t -\varphi_t \sigma^2_t d\Tilde{N}_t, \,\,\, 
 \end{equation}
with $V_0=x$, where $f$ is a \textit{nonlinear admissible driver} independent on $k$, which incorporates the imperfections in the market  and which satisfies $f(t,0,0)=0$. In the standard case of a \textit{linear} market, the driver $f$ is given by $f(t,\omega,y,z)=-yr_t(\omega)-z\theta_t$, with $\theta_t=\frac{\mu_t-r_t}{\sigma^1_t}$ (see e.g. \cite{DQS7}).

Using a change of variable which associates to $\varphi \in \mathbf{H}^2$ another process $Z \in \mathbf{H}^2$ given by $Z=\varphi \sigma^1$, one can write $\eqref{weaun}$ as follows:
\begin{equation}\label{weaun1}
-dV_t= f(t,V_t, Z_t) dt - Z_t dW_t -Z_t (\sigma_t^1)^{-1}\sigma_t^2 d\Tilde{N}_t. \,\,\, 
 \end{equation}
It can be easily observed that the market is \textit{incomplete}, as it is not possible for all $\zeta \in \textbf{L}^2(\mathscr{F}_T)$ to find $(V,Z) \in \mathbf{S}^2 \times \mathbf{H}^2$ satisfying $\eqref{weaun1}$ with $V_T=\zeta$.\\

\textbf{The set of probability measures $\mathcal{M}_0$.} Let $\mathbb{Q}$ be an equivalent probability measure to the reference probability $P$. By the martingale representation, its density process $(L_t)$ corresponds to the unique strong solution of the SDE:
\begin{align}
    dL_t=L_{t^-}(\nu_t^1dW_t+\nu_t^2d\Tilde{N}_t),
\end{align}
with $\nu^1_\cdot$ and $\nu_\cdot^2$ predictable processes such that $\nu^2_\cdot>-1$. By the Girsanov's theorem, we obtain that $W_t^{\mathbb{Q}}:=W_t-\int_0^T\nu_s^1ds$ is a $\mathbb{Q}$-Brownian motion and $\Tilde{N}_t^{\mathbb{Q}}:=\Tilde{N}_t-\int_0^T\lambda \nu_s^2ds$ is a martingale under $\mathbb{Q}$. The spaces $\textbf{S}^2_{\mathbb{Q}}$, $\textbf{H}^2_{\mathbb{Q}}$, $\textbf{L}^2_{\mathbb{Q}}(\mathscr{F}_T)$ are defined similarly to $\textbf{S}^2$, $\textbf{H}^2$, $\textbf{L}^2(\mathscr{F}_T)$, but under the probability measure $\mathbb{Q}$. The nonlinear operators $\mathscr{E}^{\mathbb{Q},g}$ and $\mathscr{Y}^{\mathbb{Q},g, \xi}$ are defined analogously to $\mathscr{E}^{g}$ and $\mathscr{Y}^{g}$, the BSDEs and Reflected BSDEs involved in the definitions being considered under the probability measure $\mathbb{Q}$. Finally, the notions of $\mathscr{E}^{g,\mathbb{Q}}$-martingale and $\mathscr{Y}^{\mathbb{Q},g,\xi}$-submartingale are similarly introduced to the ones of $\mathscr{E}^{g}$-martingale and $\mathscr{Y}^{g,\xi}$-submartingale. Whenever we use the notation $\mathscr{E}^{g}$ or $\mathscr{Y}^{g,\xi}$, it has to be understood under the reference measure $P$. \\

We remind here the notion of $f$-martingale measure (first used in a default model setting in \cite{GQS1}).
\begin{definition}
A probability measure $\mathbb{Q}$ equivalent to $P$ is a called a $f$-martingale measure if for all $x \in \R$ and $\varphi \in \textbf{H}^2 \cap \textbf{H}^2_{\mathbb{Q}}$, the wealth process $V^{x,\varphi}$ is a $\mathscr{E}_{\mathbb{Q}}^f$-martingale.
\end{definition}

Let $\mathcal{M}_0$ be the set of $f$-martingale measures such that the coefficients $\nu^1$ and $\nu^2$ are bounded. Let $\mathcal{A}$ be the set of bounded predictable processes $\alpha$ such that $\alpha>-1$.
Using the same arguments as in Proposition 3.11 in \cite{GQS1}, we derive the following characterization of the set $\mathcal{M}_0.$
\begin{proposition}
We have $\mathcal{M}_0=\left\{\mathbb{Q}^\alpha,\,\,\alpha \in \mathcal{A}\right\}$, where $\mathbb{Q}^\alpha$ admits $L^\alpha_T$ as density with respect to $P$ on $\mathscr{F}_T$, with $L^\alpha$ satisfying:
\begin{align}
    dL_t^\alpha=L^\alpha_{t^-}(-\lambda \alpha_t \sigma^2_t(\sigma_t^1)^{-1}dW_t+\alpha_t d\Tilde{N}_t);\, L_0^\alpha=1.
\end{align}
\end{proposition}

\section{Main results.} 

In this section, we present the main results of this paper. Our first contribution is to show that any process which satisfies the \textit{simultaneous}  $\mathscr{Y}^{\mathbb{Q}, g,\xi}$-submartingale property under all probability measures $\mathbb{Q} \in \mathcal{M}_0$ admits a \emph{nonlinear optional decomposition}. This result is completely new in the literature. Our second main result consists in an application of the optional decomposition to the pricing and hedging of American options from the buyer's perspective. In particular, it is the fundamental tool to get a dynamic  pricing-hedging duality result for the buyer's price of an American option (at any time $S$), and to obtain two infinitesimal characterizations of the buyer's price in terms of the \textbf{maximal subsolution} of some specific reflected BSDEs with constraints. 

\subsection{Nonlinear optional decompositions of \textit{simultaneous} $\mathscr{Y}^{\mathbb{Q}}$-submartingales}\label{Sec1}

\noindent Let $g$ be an admissible driver independent on $u$ and $\xi$ a strong semimartingale. In this subsection, we show that  all processes which satisfy the  \textit{simultaneous} $\mathscr{Y}^{\mathbb{Q},g,\xi}$-submartingale property, for all measures $\mathbb{Q} \in \mathcal{M}_0$, admit an \textit{optional decomposition}. Since the process $\xi$ and driver $g$ are fixed, we use the simpler notation $\mathscr{Y}^{\mathbb{Q}}$-submartingales.\\

\begin{theorem}[A \textit{nonlinear optional decomposition} of $\mathscr{Y}^{\mathbb{Q}}$-submartingales]\label{DEF2}

Let $(X_t)$ be a RCLL process belonging to $\mathbf{S}^2(\mathbb{Q})$, for all $\mathbb{Q} \in \mathcal{M}_0$. Suppose that  it is an $\mathscr{Y}^{\mathbb{Q}}$-strong submartingale for each $\mathbb{Q}$ $\in$ 
$\mathcal{M}_0$.
Then, there exists $Z \in \mathbf{H}^2$ and $k$, $k'$ $\in \mathbf{A}_{o}^2$  such that

\begin{align}
 -d  X_t  \displaystyle =  g(t,X_{t}, Z_{t})dt-Z_t\sigma_t^{-1}(\sigma^1_tdW_t+ \sigma^2_td\Tilde{N}_t)+dk_t-dk'_t; \label{dyn} \\
 dk_t \perp dk'_t;\nonumber \\
\int_0^T(X_{s^-}-\xi_{s^-})dk_{s}=0 {\rm \,\, a.s.} \label{SkoH}
\end{align}
Moreover, this decomposition is unique.
%
 \end{theorem}
 
 The proof of this result is based on Girsanov's Theorem and the \textit{optional decomposition} given in Proposition \ref{DEF2}, under the reference measure $P$, which also requires less integrability conditions. To this purpose, we introduce the set of \textit{admissible} drivers $g^\alpha$, for $\alpha \in \mathcal{A}$, which are given by:
 \begin{align}
     g^{\alpha}(t,\omega,y,z,k):=g(t,\omega,y,z)+\alpha_t(\omega) \lambda(k-\sigma^2_t(\sigma^1_t)^{-1}z).
 \end{align}
 
To simplify the notation, we denote $\mathscr{Y}^{g^{\alpha},\xi}$ by $\mathscr{Y}^{\alpha}$.
 
 We first provide a nonlinear \textit{predictable} decomposition of $\mathscr{Y}^{\alpha}$-submartingales, under the reference measure $P$.
 \begin{proposition}[A nonlinear \textit{predictable} decomposition of $\mathscr{Y}^{\alpha}$-submartingales]\label{preddecomp}
Let $(X_t) \in \mathbf{S}^2$ be a strong RCLL $\mathscr{Y}^{\alpha}$-submartingale for all $\alpha \in \mathcal{A}$. There exists an unique process  $(Z, U, A, A') \in (\mathbf{H}^2)^2  \times ({\mathbf{A}}^2)^2$ such that
\begin{equation}\label{firstbis1}
 -d   X_t  \displaystyle =  g(t, X_{t}, Z_{t})dt -Z_t dW_t-U_t d\Tilde{N}_t- d A'_t +d A_t,
 \end{equation}
 with $dA_t \perp dA'_t$ and 
\begin{align}\label{Skoro}
 \int_0^T(Y_{s^-}-\xi_{s^-})dA_{s}=0 {\rm \,\, a.s.}
\end{align}
and 
\begin{align}\label{cond1}
\textbf{1}_{\{ Y_{t^-}>\xi_{t^-}\}}(U_{t}-\sigma^{2}_{t} (\sigma_{t}^{1})^{-1}Z_{t})\geq 0, \,\, t \in [0,T],\,\,   dt \otimes dP {\rm \,\, a.s.}
\end{align}
and 
\begin{align}\label{measconst}
\text{The process } \int_0^\cdot\textbf{1}_{\{Y_{t^-}>\xi_{t^-}\}}\bigg(dA'_{t}-(U_t-\sigma^2_{t} (\sigma_t^1)^{-1}Z_{t} ) \lambda dt\bigg) \text{ is increasing} {\rm\,\, a.s.} \,\,\,\,  
\end{align}
\end{proposition}

\begin{remark}\label{continuity}
Assume that the process $\xi$ is left-upper semicontinuous. From equation $\eqref{firstbis1}$, we get, for all predictable stopping times $\tau \in \mathcal{T}_0$,
\begin{align}
\Delta A_\tau &=\textbf{1}_{X_{\tau^-}=\xi_{\tau^-}}(\Delta X_\tau)^-=\textbf{1}_{X_{\tau^-}=\xi_{\tau^-}}(X_{\tau^-}-X_\tau)^+ \nonumber \\
              & =\textbf{1}_{X_{\tau^-}=\xi_{\tau^-}}(\xi_{\tau^-}-X_\tau)^+ \leq \textbf{1}_{X_{\tau^-}=\xi_{\tau^-}}(\xi_{\tau}-X_\tau)^+ \leq 0\,\, \mathbb{P}-\text{a.s.}
\end{align}
Therefore, we deduce that the process $A$ is continuous.
\end{remark}


Using Proposition \ref{preddecomp}, we can provide a nonlinear optional decomposition of right-continuous left limited processes which satisfy the $\mathscr{Y}^\alpha$-submartingale property, for each $\alpha \in \mathcal{A}$.

\begin{theorem}[A nonlinear \textit{optional} decomposition of $\mathscr{Y}^{\alpha}$-submartingales]\label{DEF2}

Let $(X_t)$ be a RCLL process belonging to $\mathbf{S}^2$. Suppose that  it is an $\mathscr{Y}^{\alpha}$-strong submartingale for each $\alpha$ $\in$ ${\mathcal{A}}$.
Then, there exists $Z \in \mathbf{H}^2$ and $k$, $k'$ $\in \mathbf{A}_{o}^2$  such that

\begin{align}
 -d  X_t  \displaystyle =  g(t,X_{t}, Z_{t})dt-Z_t\sigma_t^{-1}(\sigma^1_tdW_t+ \sigma^2_td\Tilde{N}_t)+dk_t-dk'_t; \label{dyn1} \\
 dk_t \perp dk'_t;\nonumber \\
\int_0^T(X_{s^-}-\xi_{s^-})dk_{s}=0 {\rm \,\, a.s.} \label{SkoH}
\end{align}
Moreover, this decomposition is unique.
%
 \end{theorem}

\subsection{\textit{Infinitesimal characterizations} of the buyer's price of American options in an incomplete market}
Our contribution in this part consists in providing a pricing hedging duality result \textbf{at any time $S\in \mathcal{T}_0$} from the perspective of the buyer of an American option's perspective, and, using the results developed in the previous section, two  \textbf{infinitesimal characterizations of the buyer's price process} in terms of the \textbf{maximal subsolution} of two  different constrained reflected BSDE are obtained.\\
 
As in \cite{GQS2}, we first introduce the following assumption on the payoff process $(\xi_t)$: there exists $x \in \mathbb{R}$ and $\phi \in \mathbf{H}^2$ such that:
\begin{align}\label{ass}
|\xi_t| \leq V_t^{0,x}:=x-\int_0^tf(s,V_s^{0,x}, \varphi_s \sigma^1_s)ds+\int_0^t\varphi_s \sigma^1_sdW_s+\int_0^t\varphi_s \sigma^2_sd\Tilde{N}_s.
\end{align}

To define the buyer's price of the American option at each stopping time $S\in \mathcal{T}_0$, we introduce for each initial wealth $X \in L^2(\mathscr{F}_S)$, a \textit{super-hedge} against the American option from the buyer's point of view as a portfolio strategy $\varphi \in \mathbf{H}^2$ and a stopping time $\tau \in \mathcal{T}_S$ such that $V_\tau^{S,-X,\varphi}+\xi_\tau \geq 0$ a.s., where $V^{S,-X,\varphi}$ represents the wealth process associated with initial time $S$ and initial condition $X$. The \textit{buyer's price at time} $S$ is defined by the random variable 
$$\textbf{v}(S)= \esssup \{X \in L^2(\mathscr{F}_S), \,\, \exists (\varphi, \tau) \in \mathscr{B}_S(X)\},$$
with $\mathscr{B}_S(X)$ the set of all super-hedges associated with initial time $S$ and initial wealth $X$.\\ 
We introduce the driver $\bar{f}(t,\omega,y,z):=-f(t,\omega,-y,-z)$, which is clearly \textit{admissible} and denote by $\bar{\mathscr{E}}$ the associated nonlinear conditional expectation, respectively $\bar{\mathscr{Y}}$ the nonlinear operator associated to reflected BSDE with driver $\bar{f}$ and obstacle $\xi$.

We first introduce the following definition.
\begin{definition}[\textit{Predictable} reflected BSDE with constraints]\label{defbuy}
A process  $(X_t) \in \mathbf{S}^2$ is called {\rm a subsolution} of the reflected BSDE associated with driver $\bar{f}$ and obstacle $\xi$ if there exist processes $(Z,U,A, A') \in \mathbf{H}^2 \times ({\mathbf{A}}^2)^2$ such that
\begin{equation}\label{tris11}
 -d X_t  \displaystyle =  \bar{f}(t, X_{t}, Z_{t})dt-Z_tdW_t-U_td\Tilde{N}_t+dA_t-dA'_t,
\end{equation}
  with $dA_t \perp dA'_t$ and 
\begin{align}\label{Skoro11}
 \int_0^T(Y_{s^-}-\xi_{s^-})dA_{s}=0 {\rm \,\, a.s.}
\end{align}
and 
\begin{align}\label{cond11}
\textbf{1}_{\{ Y_{t^-}>\xi_{t^-}\}}(U_{t}-\sigma^{2}_{t} (\sigma_{t}^{1})^{-1}Z_{t})\geq 0, \,\, t \in [0,T],\,\,   dt \otimes dP {\rm \,\, a.s.}
\end{align}
and 
\begin{align}\label{measconst1}
\text{The process }\int_0^\cdot \textbf{1}_{\{Y_{t^-}>\xi_{t^-}\}}\bigg(dA'_{t}-(U_t-\sigma^2_{t} (\sigma_t^1)^{-1}Z_{t} ) \lambda dt\bigg) \text{ is increasing } {\rm\,\, a.s.} \,\,\,\,  
\end{align}
\end{definition}
The above BSDE is called predictable due to the fact that the increasing processes are \emph{predictable}.
\begin{theorem}[{\em Infinitesimal characterization I}]\label{buyamericano}
Let $(\xi_t)$ be a left-u.s.c. along stopping times semimartingale. Then there exists a right-continuous process $(\bar{\textbf{Y}}_\cdot) \in \textbf{S}^2$ such that $\textbf{v}(\theta)=\bar{\textbf{Y}}_\theta$, for all $\theta \in \mathcal{T}_0$ and the following dynamic pricing-hedging duality holds:
\begin{itemize}
    \item[(i)] The buyer's superhedging price process $(\textbf{v}_t)$ is a \textbf{subsolution} of the \textbf{reflected BSDE} from Definition \ref{defbuy}, i.e. there exists $(Z,U,A,A') \in \mathbf{H}^2 \times ({\mathbf{A}}^2)^2$ such that $(\bar{\textbf{Y}},Z,A,A')$ satisfies \eqref{tris11}, \eqref{Skoro11}, \eqref{cond11}, \eqref{measconst1}. Furthermore, it is the \textbf{maximal subsolution}, that is, if $(Y_t)$ is another subsolution, then $\textbf{v}_t \geq Y_t$ for all $t \in [0,T]$ a.s.
    \item[(ii)] Let $(\bar{Z}, \bar{k}, \bar{k}')$ be the associated processes to $\textbf{v}_t$ which appear in the representation \eqref{tris11}. The risky assets strategy $\bar{\varphi}:= -\sigma^{-1}\bar{Z}$ and the stopping time $\bar{\tau}_S:=\inf \{t \geq S: \,\, \textbf{v}_t=\xi_t\}$ is a superhedging strategy for the buyer, that is $(\bar{\tau}_S, \bar{\varphi}) \in \mathscr{B}_S(\textbf{v}(S))$.
\end{itemize}
\end{theorem}

\noindent We introduce now the definition of a subsolution of a specific \textit{optional reflected BSDE} (the increasing processes are optional). To this end, we first define the martingale: $M_t:= \int_0^t \sigma^1_sdW_s+\int_0^t\sigma^2_sd\Tilde{N}_s$.
\begin{definition}[\textit{Optional} reflected BSDE]\label{defbuy1}
A process  $(X_t) \in \mathbf{S}^2$ is called {\rm a subsolution} of the reflected BSDE driven by the martingale $M_t$ and associated with driver $\bar{f}$ and obstacle $\xi$ if there exists a process $(Z,k,k') \in \mathbf{H}^2 \times ({\mathbf{A}}_{o}^2)^2$ such that
\begin{equation}\label{tris22}
 -d X_t  \displaystyle =  \bar{f}(t, X_{t}, Z_{t})dt-Z_t(\sigma_t^1)^{-1}dM_t+dk_t-dk'_t,
\end{equation}
 with
\begin{align}\label{obstacle}
 X_t \geq \xi_t, \,\, t \in [0,T] \,\, {\rm a.s. }; \,\,\, X_T=\xi_T \,\, {\rm a.s.}
 \end{align}
\begin{align}\label{Skorobis}
\int_0^T(X_{s^-}-\xi_{s^-})dk_{s}=0, \,\,\,\,\, dk_t \perp dk'_t.
\end{align}
\end{definition}

\begin{theorem}[{\em Infinitesimal characterization II}]\label{buyamericano1}
Let $(\xi_t)$ be a left-u.s.c. along stopping times semimartingale. Then:
\begin{itemize}
    \item[(i)] The buyer's superhedging price process $(\textbf{v}_t)$ is a \textbf{subsolution} of the \textbf{reflected BSDE} from Definition \ref{defbuy1}, i.e. there exists $(Z,U,A,A') \in \mathbf{H}^2 \times ({\mathbf{A}}^2)^2$ such that $(\bar{\textbf{Y}},Z,A,A')$ satisfies \eqref{tris22}, \eqref{obstacle}, \eqref{Skorobis}. Furthermore, it is the \textbf{maximal subsolution}, that is, if $(Y_t)$ is another subsolution, then $\textbf{v}_t \geq Y_t$ for all $t \in [0,T]$ a.s.
    \item[(ii)] Let $(\bar{Z}, \bar{k}, \bar{k}')$ be the associated processes to $\textbf{v}_t$ which appear in the representation \eqref{tris22}. The risky assets strategy $\bar{\varphi}:= -\sigma^{-1}\bar{Z}$ and the stopping time $\bar{\tau}_S:=\inf \{t \geq S: \,\, \textbf{v}_t=\xi_t\}$ is a superhedging strategy for the buyer, that is $(\bar{\tau}_S, \bar{\varphi}) \in \mathscr{B}_S(\textbf{v}(S))$.
\end{itemize}
\end{theorem}

\begin{remark}
We point out that in the literature on pricing of American options in incomplete markets, neither a pricing hedging duality result at any time $S$, nor infinitesimal representations of the buyer's price process have been obtained (e.g. in the recent paper \cite{GQS2}, the only one result which has been established is a pricing hedging duality at time zero, and no infinitesimal characterization of the buyer's price process has been established). Our results fill this gap, and show that, despite the asymmetry between the seller and the buyer of an American option,  infinitesimal representation of the buyer's price process can be obtained in terms of the \textbf{maximal subsolution} of a specific constrained reflected BSDE.
\end{remark}

\section{Proofs}

\subsection{Proof of Proposition \ref{preddecomp}.}
We observe that by applying the $\mathscr{Y}^0$-Doob-Meyer decomposition of the RCLL strong $\mathscr{Y}^{0}$-  submartingale $(X_t)$, there exists an unique process $(Z, U,A,A') \in (\mathbf{H}^2)^2 \times ({\mathbf{A}}^2)^2$ such that
\begin{align}
-d  X_t  \displaystyle =  g(t, X_{t}, Z_{t})dt - Z_t dW_t- U_t d\Tilde{N}_t- dA'_t +dA_t; \nonumber \\
\int_0^T(X_{s^-}-\xi_{s^-})dA_s=0  {\rm \,\, a.s.}; \nonumber \\
dA_t \perp dA^{'}_t. \nonumber
\end{align}


Fix $\alpha \in {\mathcal{A}}$.  
Since $(X_t)$ is a RCLL strong $\mathscr{Y}^{\alpha}$-submartingale in ${\mathbf S}^2$ and using similar arguments as above, there exists an unique process $(Z^\alpha, U^\alpha,A^\alpha,A^{'\alpha}) \in  (\mathbf{H}^2)^2 \times ({\mathbf{A}}^2)^2$ such that 
\begin{align}\label{second}
-d  X_t  \displaystyle = \left( g(t, X_{t}, Z_t^{\alpha})+  (U^{\alpha}_t-\sigma^2_t (\sigma^1_t)^{-1}Z^\alpha_t) \alpha_t \lambda \right)dt - Z^\alpha_t dW_t- U^\alpha_t d\Tilde{N}_t- dA^{'\alpha}_t +dA^\alpha_t; \nonumber \\
\int_0^T(X_{s^-}-\xi_{s^-})dA^{\alpha}_s=0  {\rm \,\, a.s.}; \nonumber \\
dA^{\alpha}_t \perp dA^{'\alpha}_t. \nonumber 
 \end{align}

The uniqueness of the decompositions of a semimartingale and of a martingale lead to $Z_t=  Z^{\alpha}_{t}$ $dt\otimes dP$-a.s. and $K_t=  K^{\alpha}_{t}$ 
$dP\otimes dt$-a.s. This implies that $g(t, X_{t}, Z_{t})= g(t, X_{t}, Z^{\alpha}_{t})$ $dt\otimes dP$-a.s. Then, using the uniqueness of the finite variation part of the decomposition of the semimartingale $(X_t)$, we 
derive that 
\begin{equation}\label{anu2}
dA^{\alpha}_t-dA_t^{'\alpha} = dA_t-dA^{'}_t -  (U_t-\sigma^2_t (\sigma^1_t)^{-1}Z_t) \alpha_t {\lambda} dt.
\end{equation}

Since by   the Skorohod conditions $dA_t=dA^{\alpha}_t=0$ on $\{X_{t^-}>\xi_{t^-}\}$, we derive that
\begin{align}\label{eq2}
dA^{'\alpha}_t=dA'_t+(U_t-\sigma^2_t (\sigma^1_t)^{-1}Z_t) \alpha_t\lambda dt  {\rm \,\,on \,\,} \{X_{t^-}>\xi_{t^-}\}.
\end{align}

 We now show that this leads to $\textbf{1}_{\{X_{t^-}> \xi_{t^-} \}}(U_t-\sigma^2_t (\sigma^1_t)^{-1}Z_t)\lambda \geq 0$  $dt \otimes dP$ a.s.\, 
Consider the set $\mathcal{B}:=\{(U_t-\sigma^2_t (\sigma^1_t)^{-1}Z_t)\lambda<0, X_{t^-}> \xi_{t^-}\}$. Assume by contradiction that 
$P(\mathcal{B})>0$. For each $n \in \N$, define
$\alpha ^n := n{\bf 1} _{\mathcal{B}}$, which belongs to  $\mathcal{A}$.
From relation \eqref{eq2}, we get for $n$ large enough,
$E[\int_0^T\textbf{1}_{\{X_{t^-}>\xi_{t^-}\}}dA^{'\alpha^n}_t] = E[\int_0^T\textbf{1}_{\{X_{t^-}>\xi_{t^-}\}}dA'_t+ n  \int_0^T (U_t-(\sigma^2_t) (\sigma^1_t)^{-1}Z_t)\lambda  {\bf 1} _{\mathcal{B}} dt]  <0$. This leads to a contradiction, which implies that $\textbf{1}_{\{X_{t^-}> \xi_{t^-} \}}(U_t-\sigma^2_t (\sigma^1_t)^{-1}Z_t)\lambda \geq 0$  $dt \otimes dP$ a.s. We now show that $\eqref{measconst}$ holds. Assume by contradiction that there exists $\varepsilon>0$, $u,v \in [0,T]$ with $u<v$ and $D \in \mathscr{F}_T$ with $\mathbb{P}(D)>0$ such that $\int_{u}^{v} \textbf{1}_{X_{t^-}>\xi_{t^-}}\bigg(dA'_t-(U_t-\sigma^2_t (\sigma^1_t)^{-1}Z_t)\lambda dt\bigg) \leq -\varepsilon$ a.s. on $D$. Considering the sequence of controls $\alpha^n \equiv -1+\frac{1}{n}$ (which are clearly admissible) and using $\eqref{eq2}$, we get $-\frac{1}{n} \int_u^v \textbf{1}_{X_{t^-}>\xi_{t^-}}(U_t-\sigma^2_t (\sigma^1_t)^{-1}Z_t) \lambda dt \leq - \varepsilon$ on $D$. Letting $n$ tend to infinity, we get a contradiction and thus conclude that $\eqref{measconst}$ holds.

\subsection{Proof of Theorem \ref{DEF2}.}

\textit{Step 1: Existence of the decomposition} By Proposition \ref{preddecomp}, there exists an unique process  $(Z, K,A,A') \in (\mathbf{H}^2)^2  \times ({\mathbf{A}}^2)^2$ such that
\eqref{firstbis1}, \eqref{Skoro}, \eqref{cond1}, \eqref{measconst} hold. By classical results, the finite variational optional RCLL process $f_t:=A_t-A'_t-\int_0^t(K_s-\beta_s \sigma_s^{-1}Z_s)d\Tilde{N}_s $ can be uniquely decomposed as $f_\cdot=k_\cdot-k'_\cdot$, where $(k_t)$ and $(k'_t)$ are two  processes in $\mathbf{A}_0^2$ with $k_0=k'_0=0$ and $E[k_T^{'2}]< \infty$ (resp. $E[k_T^2]< \infty$), satisfying $dk_t \perp dk'_t$. Using results from Measure Theory, the measure $dk_t$ (resp. $dk'_t$) is the positive (resp. negative) variation of the measure $df_t$.
By  a slight abuse of notation, we can write:
$$ dk_t=\left(dA_t-dA'_t-(U_t-\sigma^2_t (\sigma^1_t)^{-1}Z_t )d\Tilde{N}_t\right)^{+}$$ and $$dk'_t=\left(dA_t-dA'_t-(U_t-\sigma^2_t (\sigma^1_t)^{-1}Z_t )d\Tilde{N}_t\right)^{-}.$$ 
Since $d\Tilde{N}_t= dN_t - \lambda dt$, we have
 \begin{equation}\label{decomposition-hB}
 dk_t=\bigg(dA_t-(U_t-\sigma^2_t (\sigma^1_t)^{-1}Z_t)dN_t -(dA'_t-(U_t-\sigma^2_t (\sigma^1_t)^{-1}Z_t )\lambda dt)\bigg)^{+}. 
 \end{equation}
Using the constraints \eqref{Skoro}, \eqref{cond1}, \eqref{measconst}, we derive that $\int_0^T\textbf{1}_{\{X_{t^-}>\xi_{t^-}\}}dk_t=0$. Hence, the Skorohod condition $\eqref{SkoH}$ holds. By \eqref{firstbis1} and using the definition of $f_\cdot$, we derive that equation $\eqref{dyn1}$ is satisfied.\\

\textit{Step 2: Uniqueness of the decomposition} We now show that the decomposition is unique. Let $(T_n)_{n \geq 1}$ be the sequence of jump times of the Poisson process $N$. By equation $\eqref{dyn1}$, for all $n \geq 1$, we have 
 \begin{equation}\label{sautY}
 \Delta X_{T_n}=Z_{T_n} \sigma^{-1}_{T_n} \beta_{T_n}-\Delta k_{T_n}+\Delta k'_{T_n}.
 \end{equation}
Set $B_t:=k_t-\sum_{n \geq 1} \Delta k_{T_n} \textbf{1}_{t \geq T_n}$, $B'_t:=k'_t-\sum_{n \geq 1} \Delta k'_{T_n} \textbf{1}_{t \geq T_n}$ and $X'_t:=X_t-\sum_{n \geq 1}\Delta X_{T_n}\textbf{1}_{t \geq T_n}$. Note that the non decreasing processes $B$ and $B'$ have only predictable jumps, which implies that $B, B' \in {\mathbf{A}}^2.$ Moreover, $dB_t \perp dB'_t$.
By $\eqref{dyn1}$, using $dN_t=d\Tilde{N}_t+\lambda dt$, we derive that
\begin{equation}\label{decomposition-optionnelleB4}
 -d  X'_t \displaystyle =  f(t,X_{t}, Z_{t})dt-Z_tdW_t+Z_t(\sigma_t^1)^{-1} \sigma^2_t\lambda dt+dB_t-dB'_t.
\end{equation}
By uniqueness of the semimartingale and martingale decompositions, we derive the uniqueness of the processes $Z$, $B$ and $B'$. By $\eqref{sautY}$, we get $\Delta k'_{T_n}-\Delta k_{T_n}=\Delta X_{T_n}-Z_{T_n} (\sigma^1_{T_n})^{-1} \sigma^2_{T_n}$. Since moreover $dk\perp dk'$, we finally derive the uniqueness of $k_\cdot$ and $k'_\cdot$.

\subsection{Proof of Theorem \ref {buyamericano}.}

To show Theorem \ref{buyamericano}, we first introduce a control-stopping game problem and provide several results on its associated value family.

To this purpose, we consider here the family of drivers $\{ \bar{f}^\alpha,\, \alpha \in \mathcal{A}\}$ with $\bar{f}^\alpha(t,y,z,u):=\bar{f}(t,y,z)+\alpha_t\lambda(u-\sigma_t^2(\sigma_t^1)^{-1}z)$ and the associated operators $\bar{\mathscr{E}}^\alpha$ and  $\bar{\mathscr{Y}}^\alpha$.\\

\textbf{Control-stopping game problem.} For each $S \in \mathcal{T}_0$, we define the $\mathscr{F}_S$-measurable random variable $\bar{\textbf{Y}}(S)$ as follows:
$$\bar{\textbf{Y}}(S):=  \underset{ \alpha \in {\mathcal{A}}} \essinf \,\, \underset{\tau \in \mathcal{T}_S} \esssup\,\,\bar{\mathscr{E}}^{\alpha}_{S,\tau} (\xi_\tau).$$
Note that for each $S \in \mathcal{T}_0$, $\tau \in \mathcal{T}_S$ and $\alpha \in \mathcal{A}$, $\bar{\mathscr{E}}^{\alpha}_{S, \tau}(\xi_\tau)$ depends on the control $\alpha$ only through the values of $\alpha$ on the interval $[S, \tau]$. For each $S \in \mathcal{A}$, define $\mathcal{A}_S$ the set of bounded predictable processes $\alpha$ defined on $[S,T] $ such that $\alpha_t >-1$ $dP \otimes dt$.  Therefore, we have
\begin{align}\label{value1}
\bar{\textbf{Y}}(S):=  \underset{ \alpha \in {\mathcal{A}}_S} \essinf \,\, \underset{\tau \in \mathcal{T}_S} \esssup\,\,\bar{\mathscr{E}}^{\alpha}_{S,\tau} (\xi_\tau) {\rm \,\, a.s.},
\end{align}
which, using the definition of the operator $\bar{\mathscr{Y}^{\alpha}}$, it can be written
\begin{align}\label{value1}
\bar{\textbf{Y}}(S):=  \underset{ \alpha \in {\mathcal{A}}_S} \essinf \,\, \bar{\mathscr{Y}}^{\alpha}_{S,\tau} (\xi_\tau) {\rm \,\, a.s.}
\end{align}

Under the assumption \eqref{ass}, it can be easily shown that $\mathbb{E}[\esssup_{\tau \in \mathcal{T}_0}\bar{\textbf{Y}}^2(\tau)]<\infty$.\\

We now obtain the following characterization of the family $(\overline{\textbf{Y}}(S))$.
\begin{theorem}[{\em Characterization of the family $(\overline{\textbf{Y}}(\theta))$}]\label{TH1}
We have the following characterization of the family $(\overline{\textbf{Y}}(\theta))$:
\begin{itemize}
\item[(i)] There exists a  right-continuous left-limited process $\bar{\textbf{Y}} \in \textbf{S}^2$, such that for all $\theta \in \mathcal{T}_0$, we have $\bar{\textbf{Y}}(\theta)=\bar{\textbf{Y}}_\theta$, for all $\theta \in \mathcal{T}_0$. Moreover, it is the greatest process which is a $\bar{\mathscr{Y}}^\alpha$-submartingale, for all $\alpha \in \mathcal{A}$ and it is equal to $\xi$ at the terminal time $T$.
\item[(ii)] The process $(\bar{\textbf{Y}}_t)$ is a \textit{subsolution} of the reflected BSDE from Definition \ref{defbuy}, i.e. there exists $(Z,U,A,A') \in \mathbf{H}^2 \times ({\mathbf{A}}^2)^2$ such that $(\bar{\textbf{Y}},Z,A,A')$ satisfies \eqref{tris11}, \eqref{Skoro11}, \eqref{cond11}, \eqref{measconst1}. Furthermore, it is the \textit{maximal subsolution}, that is, if $(Y_t)$ is another subsolution, then $\bar{\textbf{Y}}_t \geq Y_t$ for all $t \in [0,T]$ a.s.
\item[(iii)] The process $(\bar{\textbf{Y}}_t)$ is a \textit{subsolution} of the reflected BSDE from Definition \ref{defbuy1}, i.e. there exists $(Z,k,k') \in \mathbf{H}^2 \times ({\mathbf{A}}_{o}^2)^2$ such that $(\bar{\textbf{Y}},Z,k,k')$ satisfies \eqref{tris22}, \eqref{obstacle}, \eqref{Skorobis}. Furthermore, it is the \textit{maximal subsolution}, that is, if $(Y_t)$ is another subsolution, then $\bar{\textbf{Y}}_t \geq Y_t$ for all $t \in [0,T]$ a.s.
\end{itemize}
\end{theorem}
\begin{remark}
We point out that, within a default model in \cite{GQS2}, the existence of a process $(\bar{\textbf{Y}}_t)$ which aggregates the family $(\bar{\textbf{Y}}(S))_{S \in \mathcal{T}_0}$ has been provided using different techniques. Our method relies on the $\mathscr{Y}^{g,\xi}$-submartingales tool introduced in \cite{DESZ} and allows to characterize  the family (and associated process) as the maximal $\bar{\mathscr{Y}}^\alpha$-submartingale family (resp. greatest $\bar{\mathscr{Y}}^\alpha$-submartingale process), equal to $\xi$ at terminal time $T$, property which is further used to obtain the right-continuity of the aggregating process, and the representations in terms of maximal subsolution of reflected BSDEs (which is completely new compared to \cite{GQS2}).
\end{remark}

\paragraph{\textbf{Proof.}}

\textit{(i)}. The proof is divided in the several steps.\\

\textit{\underline{Step 1:} Existence of an optimizing sequence}. The existence of a sequence of controls $(\alpha^n)_{n \in \mathbb{N}}$ in $ \mathcal{A}_{S}$, for all $n$,  such that the sequence $(
 \mathcal{Y}^{\alpha^n}_{S,T} (\xi_{T}))_{n \in \mathbb{N}}$ is non increasing and satisfies:
\begin{equation}\label{optimizing}
\bar{\textbf{Y}}(S) \quad = \lim_{n \to \infty} \downarrow \mathscr{Y}^{\nu^n}_{S,T} (\xi_{T}) \quad
\mbox{\rm a.s.}
\end{equation}
follows by standard arguments, i.e. 
the family  $\{{\mathscr{Y}}^{\alpha}_{S,T} (\xi_T), \; \alpha \in \mathcal{A}_{S} \}$ 
is directed downward (see e.g. Proposition 7.3 in \cite{GQS2}).\\
 
\noindent \textit{\underline{Step 2}: Characterization of the family $\overline{\textbf{Y}}(S)$}. 
This step consists in showing that the family $(\bar{\textbf{Y}}(S))$ is the \textit{greatest family} such that for each $\alpha\in \mathcal{A}$, is an $\mathscr{Y}^{\alpha}$-submartingale family equal to  $\xi_T$ at terminal time $T$.
\begin{itemize}
    \item[(i)] We first show that $(\bar{\textbf{Y}}(S))$ is \textit{a $\mathscr{Y}^\alpha$-submartingale family}, for all $\alpha \in \mathcal{A}$. Let $\theta' \in \mathcal{T}_0$ and $\theta \in \mathcal{T}_{\theta'}$.  By \textit{Step 1}, there  exists $(\alpha^n)_{n \in \mathbb{N}}$ such that equality \eqref{optimizing} holds with $S=\theta$. First, notice that $\bar{\textbf{Y}}(\theta) \geq \xi_\theta$ a.s for all $\theta \in \mathcal{T}_0$. By the continuity property of reflected BSDEs with respect to terminal condition, 
$\displaystyle{
\mathscr{Y}^{ \alpha}_{\theta',\theta} (\bar{\textbf{Y}}(\theta))    =  \lim_{n\to\infty} \mathscr{Y}^{ \alpha}_{\theta',\theta}(\mathscr{Y}^{\alpha^n}_{\theta,T} (\xi_T))\,
}$ a.s. For each $n$, we set $\tilde \alpha^n_t:= \alpha_t {\bf 1}_{[\theta', \theta]}(t)+  \alpha^n_t {\bf 1}_{[\theta, T]}(t)$.
Note that $\tilde \alpha^n \in \mathcal{A}_{\theta'}$ and that $\bar{f}^{\tilde \alpha^n} = \bar{f}^{\alpha}{\bf 1}_{[\theta', \theta]} 
+ \bar{f}^{\alpha^n}{\bf 1}_{[\theta, T]}$. We thus obtain, from the consistency property of the operator $\mathscr{Y}^{ \tilde \alpha^n},$ $$\mathscr{Y}^{ \alpha}_{\theta',\theta}(\mathscr{Y}^{\alpha^n}_{\theta,T} (\xi_{T}))= \mathscr{Y}^{ \tilde \alpha^n}_{\theta',\theta}(\mathscr{Y}^{\tilde \alpha^n}_{S,T} (\xi_{T}))=  \mathscr{Y}^{ \tilde \alpha^n}_{\theta',T}(\xi_{T}) \quad {\rm a.s.}$$ 
We thus get that 
$\displaystyle{
\mathscr{Y}^{ \alpha}_{\theta',\theta} (\bar{\textbf{Y}}(\theta))   =  \lim_{n\to\infty} {\mathscr Y}^{ \tilde \alpha^n}_{\theta',T}(\xi_{T}) \,\geq \, \bar{\textbf{Y}}(\theta')}$ a.s.\,, where the last equality follows from the
 definition of $\bar{\textbf{Y}}(\theta')$.
\item[(ii)] We now show the second assertion. Let $({Y}^\prime (S),S\in \mathcal{T}_0)$ be an admissible family such that for each $\alpha \in \mathcal{A}$, it is an 
${\mathscr Y}^{\alpha}$-submartingale family such that ${Y}^\prime (T) =\xi_T$ a.s.
Let $\alpha \in \mathcal{A}$. For all $\theta \in \mathcal{T}_0$, 
$Y^\prime(\theta)\leq {\mathscr Y}^{ \alpha}_{\theta,T}({Y}^\prime(T))= {\mathscr Y}^{ \alpha}_{\theta,T}(\xi_T)$ a.s. 
Taking the essential infimum over  $\alpha\in \mathcal{A}$, we derive $Y^\prime(\theta)\leq \bar{\textbf{Y}}(\theta)$ a.s.
\end{itemize}
\noindent \textit{\underline{Step 3}: Existence of the process $\overline{\textbf{Y}}_S$}. By Theorem 2.6 in \cite{DESZ}, there exists a process $\overline{\textbf{Y}}_S=\overline{\textbf{Y}}(S)$, for all $S \in \mathcal{T}_0$, which is right-lower semicontinuous. Furthermore, by \textit{Step 2}, $\overline{\textbf{Y}}_S$ is the greatest process such that it is a $\mathscr{Y}^\alpha$-submartingale equal to $\xi_T$ at time $T$, for all $\alpha \in \mathcal{A}$. Since $(\overline{\textbf{Y}}_{t})$ is a strong $\mathscr{Y}^\alpha$-submartingale, it has left and right limits
(see Remark 2.2 in \cite{DESZ}). We define $\textbf{Y}_{t^+}:=\lim_{s \rightarrow t; s>t}\textbf{Y}_{s}$ for $0\leq t <T$ and $\textbf{Y}_{T^+}=\xi$ a.s.\\
\noindent \textit{\underline{Step 4}: The process $\overline{\textbf{Y}}_{t^+}$ is a  $\mathscr{Y}^{\alpha}$-submartingale, for all $\alpha \in \mathcal{A}$}.   
Let us first show that $(\overline{\textbf{Y}}_{t^+})$ is greater than $(\xi_t)$. Since $(\overline{\textbf{Y}}_{t})$ is a strong  $\mathscr{Y}^{\alpha}$-submartingale, by Remark 2.2. in \cite{DESZ}, it follows that $(\overline{\textbf{Y}}_{t})$ is right-l.s.c., which implies that for each $\theta \in$ ${\mathcal T}_0$, we have $\overline{\textbf{Y}}_{\theta^+}\geq \overline{\textbf{Y}}_\theta$ a.s. Since $\overline{\textbf{Y}}_\theta \geq \xi_\theta$ a.s., we derive that $\overline{\textbf{Y}}_{\theta^+} \geq \xi_\theta$ a.s.

Consider  $\theta^1, \theta^2$ $\in$ ${\mathcal T}_0$ with $\theta^1 \leq \theta^2$ a.s. 
There exist two nondecreasing sequences of 
stopping times $(\theta^1_n)$ and $(\theta^2_n)$ such that for each $n$, $\theta^1_n \leq \theta^2_n$ a.s.\,, $\theta^1_n >\theta$ a.s. on 
$\{\theta<T\}$,
$\theta^2_n > \theta$ a.s.\,on $\{\theta^2<T\}$ and $\theta^1_n \rightarrow \theta^1$ a.s. (resp. $\theta^2_n \rightarrow \theta^2$) when $n \rightarrow \infty$.
Since $(\overline{\textbf{Y}}_{t})$ is a strong 
$\mathscr{Y}^{\alpha}$-submartingale, by the consistency and the monotonicity properties of $\mathscr{Y}^{\alpha}$, we derive
$\mathscr{Y}^{\alpha}_{\theta^1, \theta^2_n} (X_{\theta^2_n})= \mathscr{Y}^{\alpha}_{\theta^1, \theta^1_n}(\mathscr{Y}^{\alpha}_{\theta^1_n, \theta^2_n} (X_{\theta^2_n}))
 \geq \mathscr{Y}^{\alpha}_{\theta^1,\theta^1_n}(X_{\theta^1_n}) \,\, {\rm  a.s.}$ Hence, since $(\xi_t)$ is  RCLL\,, we let $n$ tend to $+ \infty$ in the previous inequality and by the continuity property with respect to terminal time and terminal condition of reflected BSDEs,  we obtain 
$\mathscr{Y}^{\alpha}_{S, \theta} (\overline{\textbf{Y}}_{\theta^+}) \geq \mathscr{Y}^{\alpha}_{S, S}(\overline{\textbf{Y}}_{S^+}) = \overline{\textbf{Y}}_{S^+} \,\, {\rm a.s.}$ We thus conclude that the process $(\overline{\textbf{Y}}_{t^+})$ is a strong $\mathscr{Y}^{\alpha}$-submartingale.\\
\noindent \textit{\underline{Step 5}: The process $(\overline{\textbf{Y}}_t)$ is right-continuous left-limited.} Since by \textit{Step 4}, $(\overline{\textbf{Y}}_t)$ is a strong $\mathscr{Y}^{\alpha}$-submartingale for all $\alpha \in \mathcal{A}$ and by the maximality property of $(\overline{\textbf{Y}}_t)$, it follows that $\overline{\textbf{Y}}_t \geq \overline{\textbf{Y}}_{t^+}$, $0\leq t \leq T$ a.s.\, On the other hand,  $(\overline{\textbf{Y}}_t)$ is right-l.s.c. (cf. \textit{Step 3}). 
We thus conclude that $\overline{\textbf{Y}}_t = \overline{\textbf{Y}}_{t^+}$, $0\leq t \leq T$ a.s.\\

\noindent \textit{(ii)} By \textit{Step 3} and Theorem \ref{preddecomp}, we obtain that $(\overline{\textbf{Y}}_t)$ is \textit{subsolution} of the reflected BSDE \eqref{tris11}, \eqref{Skoro11}, \eqref{cond11}, \eqref{measconst1}. From \textit{Step 3}, we also obtain that $(\overline{\textbf{Y}}_t)$ is the greatest process which is a $\mathscr{Y}^\alpha$-submartingale, for all $\alpha \in \mathcal{A}$.\\
\noindent It remains to prove that $(\bar{\textbf{Y}}_t)$ is the maximal subsolution of the reflected BSDE \eqref{tris11}, \eqref{Skoro11}, \eqref{cond11}, \eqref{measconst1}. Assume that $(\tilde{Y}, \tilde{Z}, \tilde{U}, \tilde{A}, \tilde{A'})$ be a subsolution of the same  reflected BSDE. Let $\alpha \in \mathcal{A}$. 
 Therefore, we have
\begin{equation}\label{tris3}
 -d  \tilde{Y}_t  \displaystyle =  \bar{f}^\alpha(t,\tilde{Y}_{t}, \tilde{Z}_{t},\tilde{U}_t)dt-\alpha_t \lambda (\tilde{U}_t-\sigma_t^2 (\sigma_t^1)^{-1}\tilde{Z}_t)dt-\tilde{Z}_{t}dW_t-\tilde{U}_{t}d\Tilde{N}_t+d\Tilde{A}_t-d\Tilde{A}'_t,
\end{equation}
 with $Y_\cdot \geq \xi_\cdot$, $Y_T=\xi_T$  and the Skorohod condition
 \eqref{Skoro11}. We observe that $(\tilde{Y}, \tilde{Z}, \tilde{U}, \tilde{A}, \tilde{A'})$ also satisfies the dynamics 
 \begin{align}
     -d  \tilde{Y}_t \displaystyle = \bar{f}^\alpha(t,\tilde{Y}_{t}, \tilde{Z}_{t},\tilde{U}_t)dt-(1+\alpha)\lambda(\tilde{U}_t-\sigma_t^2 (\sigma_t^1)^{-1}\tilde{Z}_t)^+dt-\tilde{Z}_{t}dW_t-\tilde{U}_{t}d\Tilde{N}_t+d\Tilde{A}^\alpha_t-d\Tilde{A'}^\alpha_t.
 \end{align}
 Since the RCLL process $h_t:=A_t-\Tilde{A}_t'-\int_0^t(1+\alpha_t) \lambda (\tilde{U}_t-\sigma_t^2 (\sigma_t^1)^{-1}\tilde{Z}_t)^-dt+\int_0^t \lambda (\tilde{U}_t-\sigma_t^2 (\sigma_t^1)^{-1}\tilde{Z}_t)dt$ has finite variation, we can consider the associated measure and its Jordan decomposition into mutually singular measures, the  positive variation measure $d\Tilde{A}^\alpha_t$ and the negative variation measure $d\Tilde{A'}^\alpha_t$. By a slight abuse of notation, we define
$$d\Tilde{A}^\alpha_t:=\left(d{A}_t-(d\Tilde{A}_t'-(\tilde{U}_t-\sigma_t^2 (\sigma_t^1)^{-1}\tilde{Z}_t)\lambda)+(1+\alpha_t)\lambda(U_t-\sigma_t^2 (\sigma_t^1)^{-1}\tilde{Z}_t)^-dt  \right)^+$$
 and 
 $$d\Tilde{A'}^\alpha_t:=\left(d\Tilde{A}_t-(d\Tilde{A}_t'-(\tilde{U}_t-\sigma_t^2 (\sigma_t^1)^{-1}\tilde{Z}_t)\lambda)+(1+\alpha_t)(\tilde{U}_t-\sigma_t^2 (\sigma_t^1)^{-1}\tilde{Z}_t)^-dt  \right)^-.$$
 By definition, we obtain that $d\Tilde{A}^\alpha_t \perp d\Tilde{A'}^\alpha_t$. Furthermore, since $\alpha_t>-1$ and the constraints \eqref{Skoro11}, \eqref{cond11} and \eqref{measconst1} are satisfied, we obtain that $\int_0^T (\Tilde{Y}_{s^-}-\xi_{s^-})d\Tilde{A}^\alpha_t=0$ a.s. We then conclude that $(\Tilde{Y}_t,\Tilde{Z}_t,\Tilde{U}_t,\Tilde{A}^\alpha_t, \Tilde{A'}^\alpha_t)$ corresponds to the unique solution of the reflected BSDE with (generalized) driver $\bar{f}^\alpha(\cdot)dt-(1+\alpha_t)(\tilde{U}_t-\sigma_t^2 (\sigma_t^1)^{-1}\tilde{Z}_t)^+dt-d\Tilde{A'}^\alpha_t$.
 
 By applying the comparison theorem for reflected BSDEs, we have that for all $S,S' \in {\mathcal T}_0$ with $S \geq S^{'}$ a.s., 
$
\mathscr{Y}^{\alpha}_{S',S} (Y_S)  \geq  Y_{S^{'}}$ a.s.\, since $\mathscr{Y}^{\alpha}_{\cdot,S} (Y_S)$ is the solution of the reflected BSDE associated with driver 
$\bar{f}^\alpha$, obstacle $(\xi_t)$ and terminal condition $Y_{S}$.
Hence,
$(Y_t)$ is a strong $\mathscr{Y}^{\alpha}$-
submartingale for each $\alpha \in \mathcal{A}$.
Moreover, $Y_T = \xi_T$ a.s.  Hence, by \textit{Step 3}, we get 
$Y_t \leq \bar{\textbf{Y}}_t$, $0 \leq t \leq T$ a.s.\\

\noindent \textit{(iii)} By \textit{Step 3} and Theorem \ref{DEF2}, we obtain that $(\overline{\textbf{Y}}_t)$ is \textit{subsolution} of the reflected BSDE \eqref{tris22}. From \textit{Step 3}, we also obtain that $(\overline{\textbf{Y}}_t)$ is the greatest process which is a $\mathscr{Y}^\alpha$-submartingale, for all $\alpha \in \mathcal{A}$.\\
\noindent It remains to prove that $(\bar{\textbf{Y}}_t)$ is the maximal subsolution of the reflected BSDE \eqref{tris22}. Assume that $(Y, Z, K, k, k')$ be a subsolution of the same  reflected BSDE (cf. \eqref{tris22}). Let $\alpha \in \mathcal{A}$. Note that we have
\begin{equation}\label{tris3}
 -d  Y_t  \displaystyle =  \bar{f}^\alpha(t,Y_{t}, Z_{t},Z_t(\sigma_t^1)^{-1}\sigma_t^2)dt-Z_tdM_t+dk_t-dk'_t,
\end{equation}
 with $Y_\cdot \geq \xi_\cdot$, $Y_T=\xi_T$  and the Skorohod condition
 \eqref{Skorobis}. This implies that $(Y, Z, Z (\sigma^1)^{-1}\sigma^2,k)$ is the solution of the reflected BSDE associated with generalized driver $\bar{f}^\alpha(\cdot) dt-dk'_t$ and obstacle $(\xi_t)$. Using again the (generalized) comparison theorem for reflected BSDEs as in \textit{(ii)}, we have that for all $S,S' \in {\mathcal T}_0$ with $S \geq S^{'}$ a.s., 
$
\mathscr{Y}^{\alpha}_{S',S} (Y_S)  \geq  Y_{S^{'}}$ a.s.\
Hence,
$(Y_t)$ is a strong $\mathscr{Y}^{\alpha}$-
submartingale for each $\alpha \in \mathcal{A}$.
Moreover, $Y_T = \xi_T$ a.s.  Hence, by \textit{(i) - Step 3}, we get 
$Y_t \leq \bar{\textbf{Y}}_t$, $0 \leq t \leq T$ a.s.\\

We are now in position to prove the main result, Theorem \ref{buyamericano}.\\

\textbf{Proof of Theorem \ref{buyamericano}.} Fix $S \in \mathcal{T}_0$. We first show that $\textbf{v}(S) =\bar{\textbf{Y}}_S$ and $(\bar{\tau}, \bar{\varphi}) \in \mathscr{B}(\bar{\textbf{Y}}_S)$. Let $\mathscr{D}_S$  be the set of initial capitals which allow the buyer to be ``super-hedged", that is $\mathscr{D}_S= \{ X \in \mathbf{L}^2(\mathscr{F}_S): \exists (\tau, \varphi) \in \mathscr{B}(X) \}.$ It follows by definition that  $\textbf{v}(S)= \esssup \mathscr{D}_S$.

We first show that $\bar{\textbf{Y}}_S \leq \textbf{v}(S)$. 
Consider the portfolio associated with the initial capital at time $S$, i.e. $-\bar{\textbf{Y}}_S$ and the strategy $\bar{\varphi}=-(\sigma^1)^{-1}\bar{Z}$. 
By \eqref{weaun},  the value of the portfolio process $(V_t^{-\bar{\textbf{Y}}_S, \bar{\varphi}})$ satisfies the following forward differential equation:
\vspace{-3mm}
\begin{align}\label{ri}
V_t^{-\bar{\textbf{Y}}_S, \bar{\varphi}}= -\bar{\textbf{Y}}_S-\int_S^t 
f(s,V_s^{-\bar{\textbf{Y}}_S, \bar{\varphi}},-\bar{Z}_s)ds- \int_S^t \bar{Z}_s dW_s-\int_S^t (\sigma_s^1)^{-1}\sigma_s^2\bar{Z}_sd\Tilde{N}_s, \,\, S \leq t \leq T.
 \end{align}
 

Moreover, since $\bar{\textbf{Y}}$ is the solution of the reflected BSDE \eqref{tris11}, it satisfies:
\begin{equation}\label{forww}
\bar{\textbf{Y}}_t=\bar{\textbf{Y}}_S-\int_S^t \bar{f}(s,\bar{\textbf{Y}}_s,\bar{Z}_s)ds+\int_S^t \bar{Z}_sdW_s-\int_S^t (\sigma_s^1)^{-1}\sigma_s^2\bar{Z}_sd\Tilde{N}_s-\bar{A}_t+\bar{A}_S+\bar{A}'_t-\bar{A}'_S, \,\, S \leq t \leq T.
\end{equation}
We have $\bar{A}=\bar{A}^{c}+\bar{A}^{d}$, where $\bar{A}^{c}$ (resp. $\bar{A}^{d}$) is the continuous (resp. discontinuous) part of $\bar{A}$.
We first show that $\bar{A}_\cdot^{c}=\bar{A}^{c}_S$ on $[S, \bar{\tau}_S]$. Now, by definition of $\bar{\tau}_S$, we have that almost surely on $[0,\bar{\tau}_S[$, $\bar{\textbf{Y}}_t>\xi_t$. By the Skorokhod condition \eqref{Skorobis}, we get that the process $\bar{A}_\cdot^c$ is equal to $\bar{A}^c_S$ on $[S,\bar{\tau}_S[$. The continuity of $\bar{A}^{c}$ implies that $\bar{A}^c=0$ a.s. on $[S, \bar{\tau}_S]$. Under the left upper-semicontinuity assumption on the process $(\xi_t)$, by Remark \ref{continuity} we derive that $\Delta \bar{A}_\tau=0$ a.s. for all predictable stopping time $\tau \in \mathcal{T}_S$.
We multiply by $(-1)$ the equation \eqref{forww} and using the definition of the driver $\bar{f}$, we derive that  the $(- \bar{\textbf{Y}}_t)$ satisfies  the following equation:
\begin{equation}\label{forww2}
-\bar{\textbf{Y}}_t=-\bar{\textbf{Y}}_S-\int_S^t f(r,-\bar{\textbf{Y}}_r,-\bar{Z}_r)dr-\int_S^t \bar{Z}_rdW_r-\int_S^t(\sigma_r^1)^{-1}\sigma_s^2\bar{Z}_rd\Tilde{N}_r-\bar{A}'_t+\bar{A}'_S, \,\, S \leq t \leq 
\bar{\tau}_S, \,\,\, {\rm a.s.}
\end{equation}
Therefore, by the comparison result for forward differential equations,
we  get  $V_t^{-\bar{\textbf{Y}}_S, \bar{\varphi}} \geq -\bar{\textbf{Y}}_t $, $S \leq t \leq \bar{\tau}_S$ a.s. By definition of the stopping time $\bar{\tau}_S$, and the right continuity of the processes  $(\bar{\textbf{Y}}_t)$  and $(\xi_t)$, we derive that $\bar{\textbf{Y}}_{\bar{\tau}} = \xi_{\bar{\tau}}$ a.s. We thus conclude that $V_{\bar{\tau}}^{-\bar{\textbf{Y}}_S, \bar{\varphi}} \geq -\xi_{\bar{\tau}}  {\rm \,\, a.s.,}$ which implies that $(\bar{\tau}, \bar{\varphi}) \in \mathscr{B}(\bar{\textbf{Y}}_S)$ and thus $\bar{\textbf{Y}}_S\leq \textbf{v}(S)$ a.s.

The proof of the converse inequality follows by standard arguments. Let $X \in \mathscr{D}_S$. By definition of $\mathscr{D}_S$, there exists $(\tau,\varphi) \in \mathscr{B}(X)$ such that 
$V^{-X, \varphi}_{\tau} \geq -\xi_\tau $ a.s. Let $\alpha \in \mathcal{A}_S$.
We derive that
$
-X =\mathscr{E}^\alpha _{S,\tau }(V^{-X, \varphi}_{\tau}) \geq 
\mathscr{E}_{S, \tau }^\alpha (-\xi_\tau)= -\bar{\mathscr{E}}_{S, \tau }^\alpha (\xi_\tau).
$
and thus we get 
$
X \leq \bar{\mathscr{E}}_{S, \tau }^\alpha (\xi_\tau)$, which implies 
  $X\leq 
\esssup_{\tau \in \mathcal{T}_S} \bar{\mathscr{E}}_{S, \tau }^\alpha (\xi_\tau).
$
By arbitrariness of $\alpha \in \mathcal{A}_S$, we get
 $$X\leq \essinf_{\alpha \in \mathcal{A}_S} \esssup_{\tau \in \mathcal{T}_S} \bar{\mathscr{E}}_{S, \tau }^\alpha (\xi_\tau)=\bar{\textbf{Y}}_S,
$$
which holds for any $X \in \mathscr{D}_S$. By taking the essential supremum over $X \in \mathscr{D}_S$, 
we get $\textbf{v}(S) \leq \bar{\textbf{Y}}(S)$. It follows  that $\textbf{v}(S) =\bar{\textbf{Y}}(S)$ and $(\bar{\tau},\bar{\varphi}) \in \mathscr{B}(\textbf{v}(S))$.

Using that $\textbf{v}(S)=\overline{\textbf{Y}}_S$ a.s. for all $S \in \mathcal{T}_0$ and using Theorem \ref{TH1}, the result follows.

\subsection{Proof of Theorem \ref{buyamericano1}}
It follows by the same arguments as in Theorem \ref{buyamericano}, combined with Theorem \ref{TH1}, item $(iii)$.

\end{document}